**Fidelia Ibekwe-SanJuan**  (Elico – University of Lyon 3, France)
**Eric SanJuan**  (LIA, University of Avignon, France)

# Knowledge Organization Research in the last two decades: 1988-2008

**Abstract**
We apply an automatic topic mapping system to records of publications in knowledge organization published between 1988-2008. The data was collected from journals publishing articles in the KO field from Web of Science database (WoS). The results showed that while topics in the first decade (1988-1997) were more traditional, the second decade (1998-2008) was marked by a more technological orientation and by the appearance of more specialized topics driven by the pervasiveness of the Web environment.

**1: Background**

The objective of this study is to map the dynamics of research in the knowledge organization (KO) field over the past two decades (1988-2008). Some previous studies have surveyed various aspects of KO research in the past (McIlwaine & Williamson 1999, Hjorland & Albrechtsen 1999, McIlwaine 2003, Lopez-Huertas 2008, Saumure & Shiri 2008, Smiraglia 2009).

McIlwaine & Williamson (1999) analyzed trends in subject analysis research between 1988 – 1998 based on an analysis of 575 publications. In a follow-up study, McIlwaine (2003) again surveyed trends in KO between 1998-2003. The data was drawn from journals and conference proceedings but a lot of the analysis provided relied on the author's knowledge of the field. Lopèz-Heurtas (2008) provided a detailed and insightful review of what she perceived as being the current research trends in KO over "the last ten years" based on data collection from the WoS. This first data was completed with a search on the LISA database and augmented by personal readings of the ISKO conference proceedings between 1998-2006. Her findings were that mainstream research in KO were reformulations of old questions which have been around for a long time (classification, thesauri). However, recasting these research questions in the framework of the web era and especially in the era of the semantic web has given them a new life. Saumure & Shiri (2008) carried out a trend survey of KO research in the pre- and post web eras, from 1966-2006. The data used in their study was gathered from the LISTA database (Library, Information Science, and Technology Abstracts). In contrast to the previous studies, the authors introduced a more technological approach in data gathering by querying bibliographic databases for records of publications. They observed that KO research remained focused throughout the period covered on mainstream topics like cataloguing, classification. However, the pre-web era was characterized more by indexing and cataloguing issues. A shift in the focus in the post-web era became noticeable with topics like metadata generation and harvesting by computers and interoperability issues.

In the above studies, the trends perceived were as a result of human analysis and interpretation, from reading the publications and relying heavily on the expertise which the authors had of the KO domain but also of related fields (LIS, NLP, Computer sciences). While the insight offered into the evolution of research concerns in KO is not to be minimized, such manual analyses are difficult to reproduce because the parameters of the methodology are not made explicit. In particular, some studies did not make explicit the criteria for data selection, the analysis method for selecting important facts, and how the synthesis of published works was arrived at. With the notable exception of Lopez-Huertas (2008 and Saumure & Shiri (2008), the other authors did not furnish details on the dataset used and how it was gathered.

Automatic techniques for data analysis and representation have been around for a long time but have rarely if ever been used by the KO community. Smiraglia's (2009) study represents an attempt to apply such techniques to the KO field. The author applied ACA (Author Co-citation Analysis) to records of papers published between 1993-2009 in the *Knowledge Organization* (KO) journal. He sought to determine a possible North American (NA) influence in KO research by contrasting the ACA map obtained from NA authors with the one obtained from non-NA authors. ACA sheds light primarily on the intellectual base of a field, i.e, past authors whose works are being cited by publishing authors but not on current publishing authors.

**2: Motivation for current work and Data collection**

Given the ever growing volume of published works, a manual synthesis of trends in any scientific field requires a superhuman effort. Data analysis and bibliometrics offer an acceptable alternative by providing methods and tools to automatically map out the key topics, authors, journals or documents in a given field. We applied our text analysis system in order to identify key research topics in KO based on a much wider selection of journals (31) and geographic coverage (world). We studied the period between 1988-2008 and focused on the publication content of publishing articles as reflected by their titles and abstracts. To the best of our knowledge, this study represents the first attempt to apply text data analysis methods, in particular natural language processing (NLP), clustering and information visualization techniques to automatically map trends in KO research. Data collection turned out to be a bottleneck issue for KO publications. While collecting records of publications in the *KO journal* and other journals publishing KO-related studies was a relatively straight forward matter, collecting the same records for the ISKO conferences was a different kettle of fish. Records of the ISKO conference proceedings are not available in raw text format nor were they indexed in a systematic way. We then had to limit our source to journals only. We collected bibliographic records of publications from Web of Science (WoS). As previous authors had observed (Saumure & Shiri 2008), identifying publications in KO comes with the problem of delimiting the sense of knowledge organization. We manually examined the list of journals obtained from our initial query and selected 31 which published papers on KO in the KO-LIS sense between 1988-2008. This list included the ancestor of the *KO* journal formerly called "*International Classification*". A total of 931 records were obtained out of which 838 came from the *KO* journal and its ancestor. The list of journals used can be found at http://fidelia1.free.fr/isko2010/data/list-journals.pdf.

**3: Analysis Methodology**

We split the corpus into two periods: 1988-1997; 1998-2008 which we will call respectively $1^{st}$ and $2^{nd}$ decade (even if the $2^{nd}$ period covers 11 years). We then fed titles and abstracts of each period into our text mining platform TermWatch. This platform includes several text processing components. We used essentially three components of the platform in this analysis: term extraction and variant identification, term clustering and information visualization. The whole process is automated. We refer the interested reader to SanJuan & Ibekwe-SanJuan (2006) for a detailed and formal presentation of TermWatch.

### 3.1 Term extraction and term variant selection

First, domain terms were extracted based on morph-syntactic rules. Second, a term variant identifier searches for relations amongst the terms. We defined three families of terminological operations that engender semantic relations between terms: orthographic, lexico-syntactic (inclusion, substitution) and semantic (synonymy). Spelling variants and synonyms are acquired by consulting WordNet. Lexico-syntactic variations refer mainly to two linguistic operations: lexical inclusion (*aka* expansion) and lexical substitution. Lexical inclusion concerns insertions or additions of modifier words in a term as in "*classification scheme /universal classification scheme*" or of head words like in "*knowledge organization / knowledge organization system*. Lexical inclusion reflects hierarchical relations between a generic term (hypernym) and its more specific variant (hyponym). Substitutions relate terms of the same length but which vary by the change of only one word, in the same position: head substitution (*knowledge organization system / knowledge organization tool*) and modifier substitution (*generic classification scheme / universal classification scheme*).

### 3.2 Term clustering

Terms were clustered based on the presence of terminological variations between them. Cluster labels were assigned automatically by the system. Using the Pajek information visualization program (Batagelj & Mryar, 2009), we generated maps from the clusters and their links. The size of the node reflects the size of the cluster. The node colours do not have any particular signification.

### 4. Maps of Knowledge Organization research over the past two decades

We analyze the clusters and the corresponding maps obtained for the two decades. Note that in each case, the mapped clusters do not represent the entire realm of generated clusters for each period. They are the ones selected for display according to some clustering and visualization parameters.

### 4.1 Research topics in the first decade: 1988-1997

Clustering the terms based on terminological variation relations yielded 75 clusters. The cluster labels were used to generate a visualization with Pajek. Figure 1 hereafter shows the image of the clusters. The image shows one major interconnected network at the center of which we find a cluster labeled "*knowledge*". The second biggest cluster in this network is "*classification*" and the 3$^{rd}$ is "knowledge *organization number*". Other main topics visible in this 1$^{st}$ decade relate to traditional KO topics: *vocabulary control* and design of *bibliographic databases* (upper left corner); *indexing* and the related issues such as *truth*, *relation*, *description*, *user* (upper center of map), *thesaurus construction and usage* (upper center); information and *text analysis* (center left); *information-documentation* and *information science* (lower right); knowledge representation and organisation (lower east), classification schemes (center and upper right). Knowledge appears as the central axis around which other more specific themes gravitate. Indexing reflects issues related to both automatic and manual methods. Subject authority control articulates issues related to controlled vocabularies and systems to implement them (bibliographic databases). *Text analysis* is another axis of research (center left) around which we have clusters related to automated text processing (scientific text analysis, discourse analysis, natural language statement). Information-documentation and information science are directly related to the central knowledge node. Knowledge organization unfolds into organisation and representation issues, knowledge management. The "*classification*" pole draws themes on specific classification schemes (Colon, LLC, Dewey, chinese classification, Universal classification scheme, IFLA-section classification, library classification, German-language OPACS). Although most of the topics portrayed on this map are mainstream KO topics, the reliance on technology being developed in connected fields - computer sciences, library and information science – on topics such as OPACS, bibliographic databases, automatic indexing, discourse analysis, information processing storage, expert system design is apparent.

**4.1 Research topics in the second decade: 1998-2008**

Clustering the terms by variation relations alone yielded 78 clusters. Figure 2 shows the newtork obtained for this period. We find again some core topics present in the 1$^{st}$ decade:

*i- classification* research is at the center of the network with a lot of connections to the other prominent research poles;

*ii- information* is still a core research and central concern;

*iii- knowledge organization* forms the second important pole organizes topics around different facets of KO (see below for a more detailed analysis);

*iv- knowledge* clearly connected to the knowledge organization pole.

The second decade sees a bigger specialization of specific topics around major poles of "*knowledge organization*" and "*knowledge*". This shows that research on different aspects of knowledge have gained prominence in this period as evidenced by clusters labeled "*knowledge transfer, knowledge perspective, knowledge flow, knowledge domain, knowledge integration, knowledge management researcher, knowledge map, knowledge recall, knowledge representation, knowledge network, knowledge organization literature*". This could help domain specialist build a taxonomy of "knowledge-related concepts" around which research is being undertaken. "*knowledge engineering, knowledge discovery*" reflect a more computational thrust and are logically connected to the information pole rather than to the knowledge pole. Theoretical research continue to be of interest to the community with the cluster "*information science*" which is linked to the cluster "*epistemological foundation*".

Although most of the topics were already there in the first decade, we observe a more technological thrust in the way in which they are addressed in the second decade. This is evident with the appearance of clusters like "*terminology database*" or "*terminology structuring*" being linked to "*classification terminology*". The presence of cluster like "*computer algorithm, knowledge engineering, knowledge map, information retrieval domain*" reinforces this more technological oriented turn of research. The right side of the map is dominated by a more user-oriented, professional or theoretical focus of research with the presence of the "*information science, library, thesaurus, knowledge system*" poles. More importantly, we observe the emergence of new topics that were not present in the first decade :

*i- metadata:* this topic is reflected by two clusters labeled "*metadata*" and "*metadata quality*". They reflect the surge in interest of designing metadata models in the semantic web era which in turn is linked to more specific topics like folksonomy, semantic interoperability.

*ii-* "*gay-lesbian classification vocabulary*" cluster reflects publications on how to build classification system and vocabularies for describing publications on homosexuality. One study cites Ellen Greenblatt's study of gay- and lesbian-related terms in the Library of Congress Subject Headings.

*iii- web* is another new topic structuring research around web-related issues such as "*web designer, web document*".

## 5. Conclusion

The trends identified in this study have been detected automatically without requiring a human effort. We will analyze in more details the points of agreement or disagreement with earlier studies, however bearing in mind that they are not directly comparable if only because the period covered by each study is different, the dataset is different as well as the methods of analysis. We aim in the future to link publishing alongside their topics so as to show which authors are working on these topics.

## References

Batagelj V., Mryar A. (2009) Pajek. Program for Large Network Analysis. Retrieved online on 15 january 2009 [http://vlado.fmf.uni-lj.si/pub/networks/pajek/].

Dahlberg Ingetraut, (2006), Knowledge organization: a new science, *Knowledge Organization,* 33, 11-19.

Hjorland B. (2003), Fundamentals of knowledge organization, *Knowledge Organization,* 2003, 30(2), 87-111.

Hjorland B., Albrechsten H. (1999), An analysis of some trends in classification research, *Knowledge Organization,* 1999, 26(3), 131-139.

McIlwaine I.C. (2003), Trends in knowledge organization, *Knowledge Organization,* 2003, 30(2), 75-86.

McIlwaine I.C., Williamson N.J. (1999), International trends in subject analysis, *Knowledge Organization,* 1999, 26(1), 23-29.

Lopez-Huertas M., (2008) Some current research questions in the field of Knowledge organization, *Knowledge Organization,* 2008, 35(2-3), 113-136.

SanJuan E., Ibekwe-SanJuan F., Textmining without document context, *Information Processing & Management,* 42(6), 1532-1552.

Saumure K., Shiri A. (2008) Knowledge organization trends in library and information studies: a preliminary comparison of the pre- and post-web eras, *Journal of Information Science,* 2008, 34(5), 651-666.

Schiffrin R., Börner K., Mapping knowledge domains, *Publication of the National Academy of Science* (PNAS), 2004, 101(1), 5183-5185.

Smiraglia R.P., (2009), Modulation and Specialization in North American Knowledge Organization: Visualizing Pioneers, in Jacob, Elin K. and Kwasnik B. (Eds.), Pioneering North American Contributions to Knowledge Organization, *North American Symposium on Knowledge Organization,* . (2009)*,* 2, 35-46.

Figure 1. KO research topics in the 1st decade: 1988-1997.

Figure 2. KO research topics in the 2nd decade: 1998-2008.